\documentclass[runningheads]{llncs}
\usepackage{graphicx}
\usepackage{soul}
\usepackage{color}
\usepackage{cite}
\usepackage[acronym,nohypertypes={acronym},shortcuts]{glossaries}
\usepackage{graphicx}
\usepackage{subfigure}
\usepackage{multirow}
\usepackage{balance}
\usepackage{amsmath,amsfonts}
\usepackage{algorithmic}
\usepackage{algorithm}
\usepackage{nicematrix}
\usepackage{makecell}
\usepackage{array}
\usepackage{textcomp}
\usepackage{stfloats}
\usepackage{url}
\usepackage{verbatim}
\usepackage{diagbox}
\usepackage{tikz}
\usepackage{adjustbox}
\usepackage{hhline}
\usepackage{comment}

\usepackage{orcidlink}
\newcommand{\orcid}[1]{\textsuperscript{\hspace{.1em}\orcidlink{#1}\hspace{.1em}}}

\hypersetup{%
    pdfborder = {0 0 0}
}

\begin{document}
\title{HoneyEVSE: An Honeypot to emulate Electric Vehicle Supply Equipments}
%
%
\author{Massimiliano Baldo\inst{1} \and
Tommaso Bianchi\inst{1}\orcid{0000-0001-8192-5117} \and
Mauro Conti\inst{1}\orcid{0000-0002-3612-1934} \and
Alessio Trevisan\inst{1} \and
Federico Turrin\inst{1}\orcid{0000-0001-5660-2447}}
\authorrunning{M. Baldo et al.}
%
\institute{University of Padova, Padua, Italy\\
\email{\{massimiliano.baldo, alessio.trevisan.2\}@studenti.unipd.it}\\
\email{\{tommaso.bianchi, federico.turrin\}@phd.unipd.it}\\
\email{mauro.conti@unipd.it}}
\maketitle              

\newacronym{ac}{AC}{Alternate Current}
\newacronym{bms}{BMS}{Battery Management System}
\newacronym{can}{CAN}{Controller Area Network}
\newacronym{cps}{CPS}{Cyber-Physical System}
\newacronym{dc}{DC}{Direct Current}
\newacronym{ecu}{ECU}{Electronic Control Unit}
\newacronym{ev}{EV}{Electric Vehicle}
\newacronym{evse}{EVSE}{Electric Vehicle Supply Equipment}
\newacronym{fpga}{FPGA}{Field Programmable Gate Array}
\newacronym{ftp}{FTP}{File Transfer Protocol}
\newacronym{gui}{GUI}{Graphical User Interface}
\newacronym{hvac}{HVAC}{Heating, Ventilation, and Air Conditioning}
\newacronym{lin}{LIN}{Local Interconnect Network}
\newacronym{mitm}{MitM}{Man-in-the-Middle}
\newacronym{most}{MOST}{Media Oriented Systems Transport}
\newacronym{tcp}{TCP}{Transmission Control Protocol}
\newacronym{tls}{TLS}{Transport Layer Security}
\newacronym{v2g}{V2G}{Vehicle-to-Grid}
\newacronym{ocpp}{OCPP}{Open Charge Point Protocol}
\newacronym{wpt}{WPT}{Wireless Power Transfer}
\newacronym{dos}{DoS}{Denial of Service}
\newacronym{soc}{SoC}{State of Charge}
\newacronym{ccs}{CCS}{Combined Charging System}
\newacronym{rfid}{RFID}{Radio-Frequency IDentification}
\newacronym{hmi}{HMI}{Human-Machine Interface}
\newacronym{iot}{IoT}{Internet of Things}
\newacronym{url}{URL}{Uniform Resource Locator}
\newacronym{mqtt}{MQTT}{MQ Telemetry Transport}

\newacronym{plc}{PLC}{Programmable Logic Controller}
\newacronym{it}{IT}{Information Technology}
\newacronym{isp}{ISP}{Internet Service Provider}
\newacronym{bgp}{BGP}{Border Gateway Protocol}
\newacronym{igp}{IGP}{Interior Gateway Protocol}
\newacronym{as}{AS}{Autonomous System}
\newacronym{tp}{TP}{Transit Provider}
\newacronym{ixp}{IXP}{Internet Exchange Point}
\newacronym{ics}{ICS}{Industrial Control Systems}

\newcommand{\tommaso}[1]{\textcolor{blue}{Tommaso: #1}}
\newcommand{\fede}[1]{\textcolor{red}{Fede: #1}}
\newcommand{\rev}[1]{\textcolor{teal}{#1}}
\newcommand{\todo}[1]{\textcolor{red}{TODO}}
\newcommand{\parag}[1]{\noindent\textbf{#1. }}
\begin{abstract}
To fight climate change, new ``green'' technology are emerging, most of them using electricity as a power source. Among the solutions, Electric Vehicles (EVs) represent a central asset in the future transport system. EVs require a complex infrastructure to enable the so-called Vehicle-to-Grid (V2G) paradigm to manage the charging process between the smart grid and the EV. In this paradigm, the Electric Vehicle Supply Equipment (EVSE), or charging station, is the end device that authenticates the vehicle and delivers the power to charge it. 
However, since an EVSE is publicly exposed and connected to the Internet, recent works show how an attacker with physical tampering and remote access can target an EVSE, exposing the security of the entire infrastructure and the final user. For this reason, it is important to develop novel strategies to secure such infrastructures.

In this paper we present HoneyEVSE, the first honeypot conceived to simulate an EVSE. HoneyEVSE can simulate with high fidelity the EV charging process and, at the same time, enables a user to interact with it through a dashboard. Furthermore, based on other charging columns exposed on the Internet, we emulate the login and device information pages to increase user engagement. 
We exposed HoneyEVSE for 30 days to the Internet to assess its capability and measured the interaction received with its Shodan Honeyscore. Results show that HoneyEVSE can successfully evade the Shodan honeyscore metric while attracting a high number of interactions on the exposed services.

\end{abstract}

\section{Introduction}
To fight climate change, novel technologies are emerging to reduce the emission footprint. Among the different promising solutions, \ac{ev} aims at substituting traditional fossil-fueled vehicles to achieve better performances and fewer emissions. According to Statista~\cite{statista_ev}, by 2027, the number of \ac{ev} sold in the market will be around 16.21m. To manage such a huge amount of vehicles is required an ad-hoc infrastructure. Indeed, to deliver the energy needed to recharge \acp{ev}, a connection between the \ac{ev} and the power grid is required. The paradigm that regulates such a connection is called \ac{v2g}. \ac{v2g} paradigm includes three main entities: 1) the smart grid, which generates and distributes electric energy; 2) the \ac{ev} that represents the end-user asking for a recharge 3) the charging column, or \ac{evse}, that authenticate the \ac{ev} and deliver the energy.

Managing a distributed and complex architecture is very challenging, including from the cybersecurity perspective.
In fact, recent works showed the feasibility of a wide range of attacks on \ac{v2g} infrastructure, particularly the \ac{evse} devices. \ac{evse} are publicly exposed and therefore subject to physical manumission by malicious actors. Attacks that have been proven effective in targeting \ac{evse} include relay attack~\cite{EVExchange}, charging traces profiling~\cite{brighente2021evscout2}, eavesdropping~\cite{baker2019losing}, and denial of service~\cite{kohler2022brokenwire}. The threats affecting \ac{evse} are even more emphasized by recent work highlighting the lack of security policies and the exposition to the Internet, opening dangerous vulnerabilities surfaces to the user~\cite{nasrchargeprint}.

\parag{Motivation} Being a relatively recent technology, the security of \ac{v2g} paradigm and, in particular, \ac{evse} devices is still under investigation, several novel attacks have been identified, and therefore research still requires contribution in this direction. To this end, honeypots can support the research of new security mechanisms. The goal of a honeypot is twofold: it can be used to deceive the attackers, making them think they are interacting with a real device and collecting data about the attackers' movements. To develop more robust defense mechanisms, data collected from honeypot can then be analyzed to understand the typical attackers' strategies and scanning campaigns. Honeypots, are widely adopted to emulate \ac{it} devices or services. However, their application in the \ac{cps} domain is still under development, mainly due to the difficulty in replicating devices and physical processes with high fidelity~\cite{conti2022icspot}. 

\parag{Contribution}
In this paper, we present HoneyEVSE, the first honeypot conceived to emulate an \ac{evse}. To build an effective honeypot with high fidelity, we based our implementation on the exposed \ac{evse} device we identified and the related work analyzing the exposure of such systems~\cite{nasrchargeprint}. To mimic a realistic charging process, we leveraged the ACN-sim simulator, while the network personality engine is managed by a customized version of Honeyd that incorporated \ac{evse} functionalities.
We exposed HoneyEVSE for 30 days on a local \ac{ixp}, and we measured different engagement results, as commands exchanged with the Web app, the time spent on each page by the user, and the origin of the IPs. Results show that HoneyEVSE can successfully evade Shodan's Hoenyscore while receiving a high number of interactions from users.
We release the source code of HoneyEVSE on github\footnote{Repository: \url{https://github.com/spritz-group/HoneyEVSE}}, and we summarize the contribution of the paper in the following:
\begin{itemize}
    \item We present HoneyEVSE, the first honeypot conceived to emulate an \ac{evse}. In HoneyEVSE, we integrate a high-fidelity physical process and the possibility for an attacker to interact with the dashboard.
    \item We exposed HoneyEVSE to the Internet for 30 days, and we report the results of the interactions received. The results highlight the good level of interaction received by HoneyEVSE.
\end{itemize}

\parag{Organization}
The paper is organized as follows.
Section~\ref{sec:background} briefly recalls the concepts useful to understand the remainder of the paper, while Section~\ref{sec:related} discusses the related work.
Section~\ref{sec:honeyevse} introduces HoneyEVSE honeypot structure and functionalities, then Section~\ref{sec:results} presents the measurement study we performed by exposing HoneyEVSE to the Internet.
Finally, Section~\ref{sec:conclusion} concludes the paper.

\section{Background}\label{sec:background}

In the following, we briefly recall the main concepts useful to understand the remainder of the paper. In particular, Section~\ref{subsec:ASIXP} recalls the \ac{ixp} infrastructure, where we based our evaluation phase. Then Section~\ref{subsec:ics_honey_back} and Section~\ref{subsec:v2g_back}
briefly recall, respectively, the concept of honeypot and \ac{v2g} infrastructure. 
\subsection{Internet Exchange Point}\label{subsec:ASIXP}

An \ac{ixp} is a network facility that enables the interconnection and Internet traffic exchange between two or more independent \acp{as} through specific peering agreements and according to the \ac{bgp} routing configurations. Its typical architecture consists of single or multiple switches connected to the adherent's border routers of the adherent \acp{as}, ensuring bandwidth, costs, and latency benefits. An \ac{as} comprises a group of IP prefixes controlled by a single \ac{isp}, which defines the routing policies. The \ac{igp} enables the routing within an \ac{as}, while communication with other \acp{as} relies on the \ac{bgp}. An \ac{ixp} network facility enables Internet traffic interconnection between more than two independent \acp{as}. 

\begin{figure}[t]
    \centering
    \includegraphics[width=0.6\columnwidth]{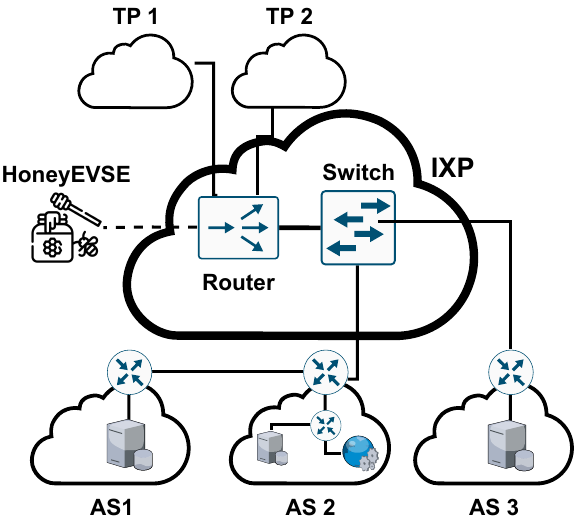}
    \caption{IXP System representation and corresponding installation point of the honeypot.}
    \label{fig:sys_model}
    \vspace{-1em}
\end{figure}

Our analysis relies on \emph{VSIX}~\cite{vsix}, an \ac{ixp} which manages the traffic circulating in the North East of Italy.
In this work, we installed a honeypot in VSIX and exposed it to the Internet. The system architecture of the \ac{ixp} representation, together with the honeypot, is represented in Figure~\ref{fig:sys_model}.
Thanks to their infrastructure VSIX allows us to integrate many interesting features to the honeypot exposed. In particular, the IP dedicated to the honeypot has been announced worldwide thanks to the transit provider, allowing high visibility worldwide.
Furthermore, VSIX offers high resilience thanks to a \ac{bgp} multihomed system and supports a connection speed of up to 1 Gb/s.

\subsection{Honeypot}\label{subsec:ics_honey_back}

Honeypots are systems designed to protect systems and, at the same time, collect information related to attacker actions. Once deployed, honeypots are exposed over the Internet and may include exploitable vulnerabilities and services. If carefully monitored, honeypots allow for obtaining a lot of information and insights about the attacks registered.

Honeypots are classified based on the level of interaction they offer to an attacker. \textit{Pure Honeypots} are real machines installed and exposed in the production network (e.g., PLC exposed but not employed in controlling processes). Although this seems to be the best approach to installing a honeypot, it also has disadvantages, most notably its cost. \textit{High-Interaction Honeypots} are generally real computers, or virtual machines, that simulate with high fidelity all the services of the emulated machine. They allow a high level of interaction with the emulated system. \textit{Low-Interaction Honeypots} are software that emulates the operating system and services provided by the simulated device. They are easy to install and maintain but, due to the lower degree of engagement it provides, make it possible to capture less information. Furthermore, honeypots can also be classified based on their scope. \textit{Production Honeypots} are usually low-interaction and simple to install and use. They are located within an enterprise production network along with production servers. Their primary function is not to collect information but to raise the alarm if detecting an attacker's presence. \textit{Research Honeypots}, contrary to the previous typology, aims at collecting as much information as possible about potential attackers. For this reason, they require a higher level of interaction.

In our work, we developed a \textit{High-Interaction} \textit{Research} honeypot specifically designed to simulate an \ac{evse}.

\subsection{Vehicle-to-Grid (V2G)}\label{subsec:v2g_back}

\begin{figure}[t]
\centering
\includegraphics[width=0.75\columnwidth]{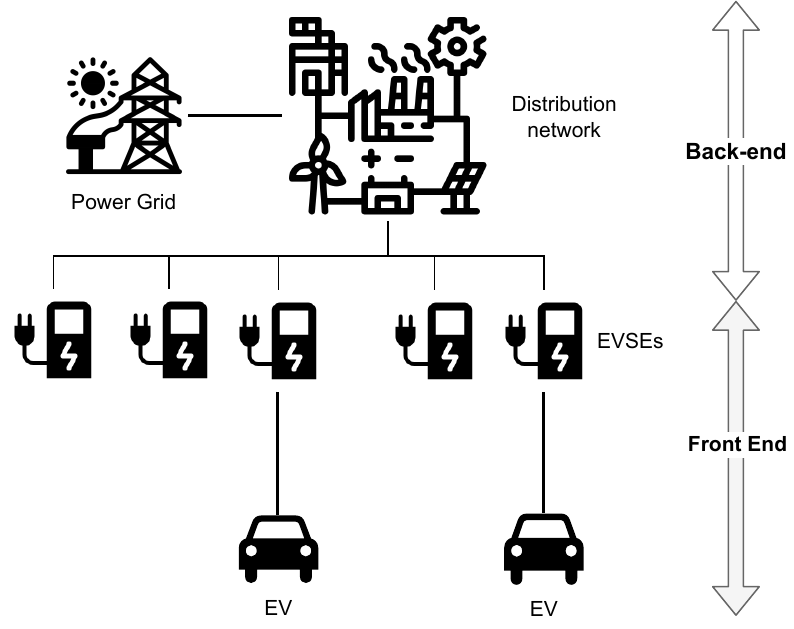}
\caption{The basic architecture of a V2G environment.}
\label{fig:v2g-architecture}
\end{figure}

The \ac{v2g} paradigm (depicted in Figure~\ref{fig:v2g-architecture}) creates a bidirectional communication between an \ac{ev} and a power grid. 
Among the different advantages, the bi-directionality of communication allows smart management of the charging process and enables the \ac{ev} to create a bidirectional power flow with the grid. 

The \ac{v2g} paradigm includes different entities: the power grid, responsible for generating electricity, the distribution network, which involves all the infrastructures to deliver the energy demanded to the final user; the \ac{evse} that authenticates the end user and deliver the energy, and the \ac{ev} that represents the end user and requires the energy delivery. To authenticate the user, the \ac{evse} is generally connected to an energy provider and accounts for the \ac{ev} energy demand. \acp{evse} include different components: charging station, charging cable, and charging connector. The last one is usually attached to the cable~\cite{evse-anatomy}. The \ac{evse} performs all the actions required to recharge the \ac{ev} and interface it with the \ac{v2g} infrastructure. Moreover, an \ac{evse} usually integrates a graphical interface to allow users to monitor and control the state of the charge of their vehicle. The technology implemented in an \ac{evse} can differ from manufacturers, but it usually relies on an HTTP service for a Web App interface, an SSH connection for remote access by an administrator, and different kinds of connectivity such as WiFi or Bluetooth~\cite{openenergymonitor}. Additionally, an \ac{evse} can support \acrfull{mqtt} connection in the case of aggregated \ac{evse}, for example, as a charging station with multiple columns.
We can divide the \ac{v2g} communication into the front-end and back-end. Back-end communication enables the exchanges of data and energy between the \ac{evse} and the distribution network. The most adopted protocol in this type of communication is the \ac{ocpp}~\cite{garofalaki2022electric}.
Similarly to back-end communication, front-end communication includes both power and data delivery. Different standards have been proposed for front-end communications to regulate the communication between the \ac{ev} and the \ac{evse}. 
The most widely adopted protocols for the front-end communication between the vehicle and the \ac{evse} are ISO 15118, SAE J2847, and CHAdeMO. Among them, the most advanced standard is ISO 15118~\cite{ISO15118-1, ISO15118-2} which supports many services, including secure authentication, vehicle firmware update, and plug-and-charge~\cite{Buschlinger2019}.

\section{Related Work}\label{sec:related}

To the best of our knowledge, HoneyEVSE is the first honeypot conceived to emulate an \ac{evse}; therefore, there is no other literature to compare with. However, honeypots are a warm topic, and significant efforts have been produced in this research direction~\cite{franco2021survey}. In the \ac{it} domain, honeypots have been used to emulate common \ac{it} services such as Databases~\cite{cenys2004development}, Web Servers~\cite{rahmatullah2016implementation}, or \ac{iot} devices~\cite{luo2017iotcandyjar}. This family of honeypots is the most commonly investigated in the literature also due to its simplicity of emulation: they are just required to replicate a service in a protected environment.
Other than the \ac{it} domain, honeypots are highly studied in the \ac{cps} domain~\cite{franco2021survey}. Particular focus has been dedicated to the Industrial domain, which has been shown to be dramatically exposed to cyberattacks in recent years~\cite {9522219}. To protect such systems, several honeypots have been proposed~\cite{HoneyPLC} ranging from water systems~\cite{murillo2018virtual, petre2019honeypot} to smart grids~\cite{gridpot_proj, mashima2019s}. The most difficult part in developing this type of honeypot is emulating a real physical process with the possibility of interacting with it by the attacker. Indeed only a few honeypots aim at addressing this challenge~\cite{conti2022icspot}. A few Honeypots have also been proposed to replicate a vehicular domain~\cite{9903648, sharma2018survey}. However, these honeypots are specifically designed to replicate  vehicular ad-hoc networks, but none consider the \ac{v2g} communication paradigm together with its specific devices.  

Among the other numerous \ac{cps} domains, the \ac{ev} field is nowadays under the spotlight. This is mainly due to the pervasive proliferation of this technology in daily life but also due to research highlighting the current security flaws of such a system. Indeed, different researchers have successfully proven attacks on \ac{v2g} infrastructure, where the entry point is mostly the \ac{evse}. These attacks include relay attack~\cite{EVExchange}, charging traces profiling~\cite{brighente2021evscout2}, eavesdropping~\cite{baker2019losing}, and denial of service~\cite{kohler2022brokenwire}. Furthermore, recent research has shown the problematic exposure of the \ac{evse} to the internet~\cite{nasrchargeprint}. Driven by this emerging threat, we present HoneyEVSE to support the security community to protect \ac{evse} and, at the same time, collect data about attackers' moves.

\section{HoneyEVSE Honeypot}\label{sec:honeyevse}

HoneyEVSE, as a honeypot, is designed to deceive attacks targeting real EVSE and retrieve information about the techniques used by attackers. It emulates the charging process of electric vehicles, exposing information through a web application accessible from the Internet. The general functioning and architecture are inspired by two previous projects: HoneyPLC~\cite{HoneyPLC} and ICSPot~\cite{conti2022icspot}.

In this Section, we first introduce HoneyEVSE. In Section~\ref{subsec:architecture}, we describe the honeypot architecture and, in Section~\ref{subsec:phy_process}, Section~\ref{subsec:services}, and Section~\ref{subsec:logs}, the technical aspects of HoneyEVSE concerning the physical process, the exposed services, and the logs, respectively. \par

\subsection{Architecture}\label{subsec:architecture}

\begin{figure}[t]
    \centering
    \includegraphics[width=0.7\columnwidth]{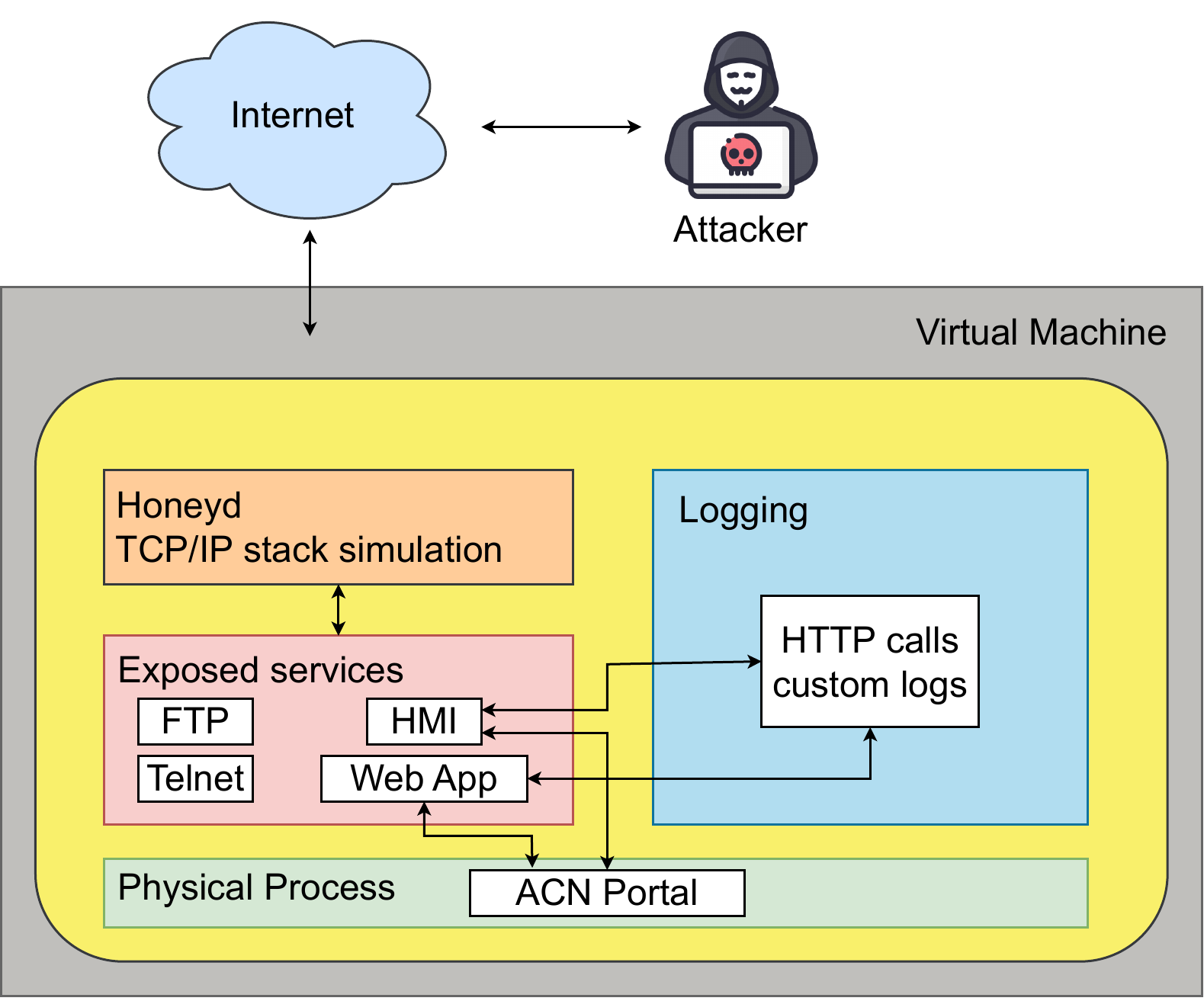}
    \caption{HoneyEVSE architecture.}
    \label{fig:honeyevse-architecture}
\end{figure}

HoneyEVSE integrates different components to generate the physical process, the network exposition, and the interaction logging. In Figure~\ref{fig:honeyevse-architecture}, we show the graphical representation of the honeypot architecture. 
The core part of HoneyEVSE is Honeyd~\cite{Honeyd}, a tool that emulates the TCP/IP network stack and defines the honeypot network profile. It allows the system to differentiate the IP and MAC addresses from the ones of the hosting computer and trigger the different components of the systems once the interaction is received. 

The only necessary step to use Honeyd is to provide, in a configuration file, the parameters for the implemented services, the IP address on which we want to route the honeypot communication, and the physical interface (MAC address) of the IP of the hosting machine. 
Honeyd module is then linked to the physical process simulating the charging activity and a logging system.

\subsection{Physical Process}\label{subsec:phy_process}

To make the honeypot as real as possible from an external perspective, we introduce a physical process emulating the vehicle's charging. The physical process leverages ACN Portal~\cite{ACNPortal}, a tool suite that allows \ac{ev} researchers to develop and test practical solutions. 
ACN Portal, accessible via public API, combines ACN-Data~\cite{lee2019acn} and ACN-Sim~\cite{ACNSim}. 
The former is a dataset containing a data collection of charging sessions of real \ac{ev}, gathered by the authors through \ac{evse} installed at the Caltech Institute of Technology.
The latter is a simulation environment for \ac{ev} charges which enables the study of algorithms for scheduling the power and the time of the charging processes. 
HoneyEVSE uses ACN Portal to simulate the underlying physical process and create reliable and truthful charging operations to deceive the attacker. 
In particular, we leveraged these tools to generate real-time EV charging traces and build a GUI with an interactive dashboard to monitor and interact with the charging process, similar to what happens in real life. 
However, the default charging traces generation of the ACN Portal presents some limitations. The representation of the \ac{ev} and the \ac{ev} is limited to a single case: a completely discharged vehicle with a single static template of charging parameters; therefore ACN-Sim does not allow for including variability in the charging process. Furthermore, ACN-Sim charging parameters do not include helpful information to replicate a real-world scenario: energy required to fill the battery, current completion percentage, and the recharge cost. Excluding this information does not allow to achieve a high fidelity level of the physical process. 
In the HoneyEVSE physical process, we address these limitations by improving the data generation and estimating the parameters from the traces generated. In particular, we associate each charging trace generated by fictional vehicles with random arrival and departure times, energy requests (e.g., the power required by the vehicle), and battery parameters (e.g., capacity). In this way, an external viewer has the feeling to interact with a real environment.
As previously said, ACN-Sim does not include the case where the initial charge of the car's battery is greater than zero. This may raise some warning in the attacker that would always see a completely discharged vehicle.
We addressed this limitation by developing a heuristic to calculate a random plausible initial status using the default information provided by ACN-Sim and considering the charging demand of the specific vehicle and its overall requested energy.

\begin{figure}[t]
    \centering
    \includegraphics[width=0.8\columnwidth]{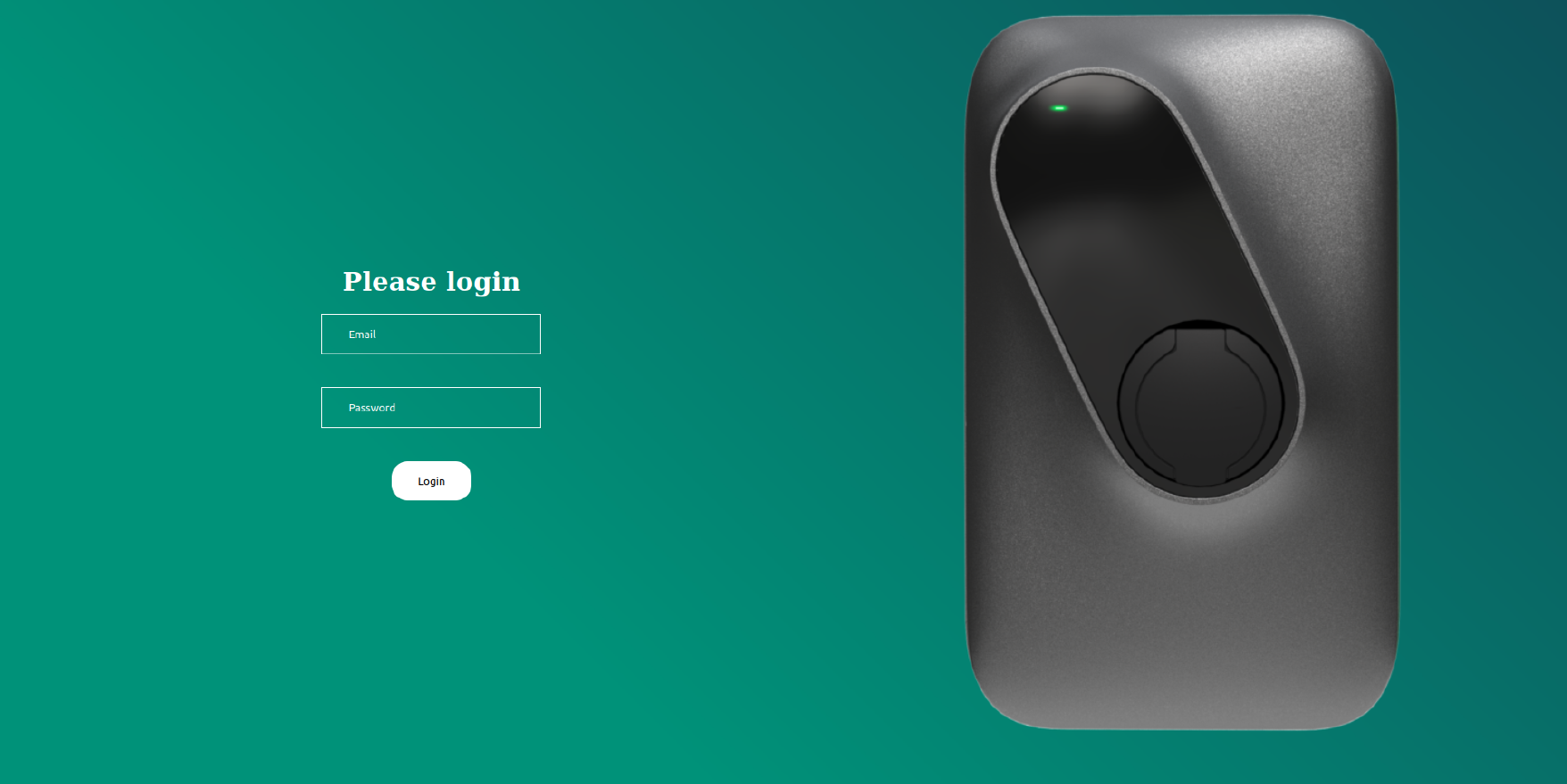}
    \caption{Login page replicated from Etrel \ac{evse}}
    \label{fig:webapp-login}
\end{figure}

\subsection{Services and Interaction}\label{subsec:services}

HoneyEVSE exposed three different services: a web interface to let the attacker interact with and the traditional Telnet and \ac{ftp} services to mimic the possibility of remotely updating or controlling charging columns. The web interface is built by taking as a reference the \ac{evse} we found in the wild using Shodan~\cite{Shodan} upon a preliminary analysis. 

\parag{HTTP} Inspired by what we and related work found exposed on the Internet, we mock a web application of a charging column. The web interface reports the data generated by the physical process, the \ac{evse} technical details of the charging column, and a login page.
On the root page, the interface shows technical information about the \ac{evse}. We copied the layout and template of such information from existing exposed \ac{evse} and modified the existing information. We implemented a dedicated dashboard with a real-time charging process to monitor the physical process from the perspective of the \ac{ev} owner that is recharging the car. This dashboard includes three buttons: Stop, Pause, and Resume Charge, and they represent the possible interaction of a user with the charging column. Furthermore, we implemented an admin dashboard with information and statistics on the different vehicles attached and the demands from the grid. This page represents the overview of the different processes occurring on the \acp{evse} that the admin generally monitors. These three pages are hosted on port 5000.
On port 80, we also implemented a \textit{login} and \textit{registration} pages to mock login and log-on procedures of users and company employees. The actions performed using these forms are logged, and every login attempts returns an error by default. We copied the login page template from Etrel \ac{evse} exposed on the Internet. Indeed, according to our preliminary analysis and related work~\cite{nasrchargeprint}, this type of charging column is widely exposed on the Internet.

\parag{FTP} The \ac{ftp} allows the file transmission over TCP/IP connections. The service is open on port 21, waiting for commands. This behavior emulates an administrator's ability to remotely perform firmware updates or web application changes. The exposed service waits for a user to log in but always retrieves an error during the login phase. Thanks to this service, we can monitor the possible attempts of an attacker to access the \ac{evse}.

\parag{Telnet} This protocol allows two-way unencrypted text-based communication between two machines. It provides access to a virtual terminal in the remote system, and it can be used to control the EVSE. This service uses port 23. As for \ac{ftp}, the system returns an error during the communication to avoid possible exploits. Meanwhile, it collects data about the attacker's attempts to access the \ac{evse} through Telnet.

\subsection{Data Logging}\label{subsec:logs}

HoneyEVSE includes a logging module connected to the Web App interface to record the interaction with the accessible services. The web app logger intercepts different communications and divides the logs into three main categories:
\begin{itemize}
    \item \textbf{Port}. We store in the logging file the number of port with wich the attacker interact;
    \item \textbf{Actions}. The physical process dashboard contains three buttons that enable a user to manipulate the charging process. The three buttons are ``Stop,'' ``Pause,'' and ``Resume'' and they include a javascript linked to them to log in when a person clicks on them. Furthermore, to simulate a user's interaction with the charging columns, the buttons trigger the corresponding actions in the physical process. 
    \item \textbf{HTTP Requests}. We log the HTTP requests directed to the honeypot. The log file includes the specific page requested, the \ac{url} that can contain a malicious payload, the request code, and the request result;
    \item \textbf{Timing}. We also report each user's time on a specific page. To do this, we use the TimeMe.js\footnote{\url{https://github.com/jasonzissman/TimeMe.js}} javascript library.
\end{itemize}

In all of these log types, we store the IP of the remote host generating the action, along with the time and date of the log.

\section{Results}\label{sec:results}

We exposed for 30 days an instance of HoneyEVSE through a virtual machine hosted by VSIX IXP. 
We verified the honeyscore assigned from Shodan to HoneyEVSE as a preliminary analysis. The honeyscore is a score from $0$ to $1$ that Shodan assigns to an IP address and corresponds to the likelihood that a specific IP is a honeypot. Ideally, the goal is to achieve a lower Honeyscore as possible. In this way, an attacker cannot understand that it is interacting with a honeypot.
Unfortunately, Shodan has not released detailed information on the computation of the honeyscore. However, in~\cite{shodan_guide}, the author states that the computation is based on a combination of metrics: (1) the number of open ports; (2) matching between services and the environment (3) honeypots default settings; (4) the IP history; (5) a not disclosed Machine Learning algorithm. Shodan could not assign a score to HoneyEVSE, meaning our honeypot can successfully camouflage to Shodan's detection.
In the following, we describe the results by means of interaction received on the web application and the IPs originating the scan. We present first in Section~\ref{subsec:res_interanal} an analysis of the interaction received by HoneyEVSE, then in Section~\ref{subsec:res_scanorigin} we present an analysis of the actors we registered.

\subsection{Interactions Analysis}\label{subsec:res_interanal}
\begin{figure}[t]
     \centering
     \includegraphics[width=0.75\columnwidth]{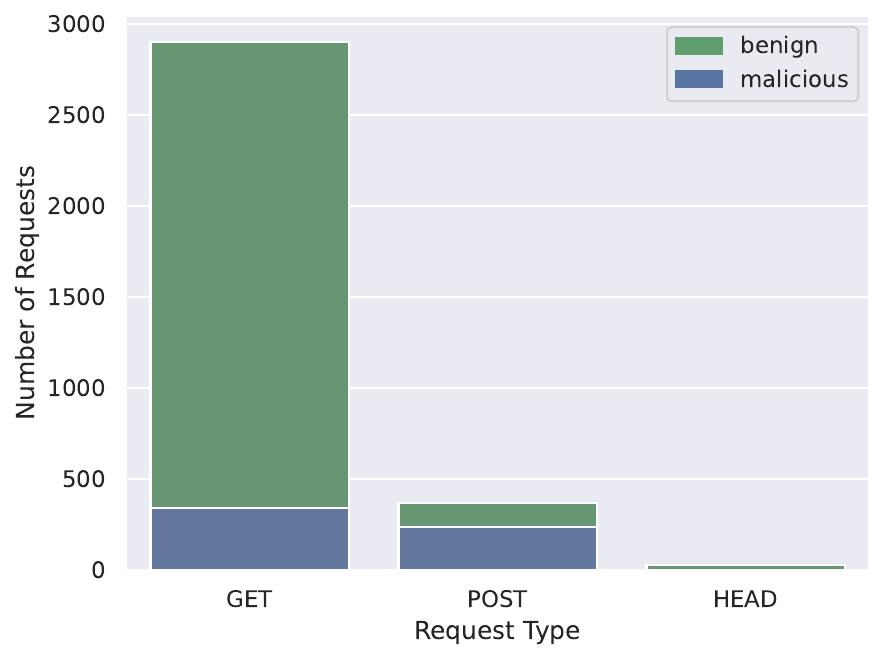}
     \caption{Type of interactions received.}
      \label{fig:request_results}
\end{figure}

After 30 days of data collection, we report a total of 3293 HTTP requests on the three different pages. Specifically, we found 2899 GET requests, 366 POST requests, and 28 HEAD requests. In the following, we analyze the type of HTTP requests received.
We define a request as malicious if it contains patterns recalling an attack behavior or attempt. For instance, the request for the URL \textit{"/shell?cd+/tmp;rm+-rf+*;wget+167.71.210.63/jaws;sh+/tmp/jaws"} seeks a remote code execution followed by a remote file upload attack. Also, different requests try to invoke \textit{cgi-bin} scripts, a common folder with programs aiming to interact with a Web browser. Lastly, many requests target the database, trying to exploit MySQL or PHP vulnerabilities, in line with the vulnerabilities \ac{evse} vulnerabilities identified in~\cite{nasrchargeprint}. Among all the requests, we classified as malicious 340 GET and 236 POST actions. Almost all of them seem to be directed by automatic tools or bots due to the absence of a specific target. However, we have not identified requests tailored specifically for HoneyEVSE charging process. Instead, all the scans and attacks were directed against the web interface. We show the results in Figure~\ref{fig:request_results}.
We have not found significant activities regarding the time spent on each page. Indeed, on average, the visitors spent about 2 seconds on each page. This may imply that they were visited through automatic navigation tools, not humans.
Finally, we have not identified particular activities on \ac{ftp} and Telnet ports, and on the login page form.

\subsection{Interactions Origin}\label{subsec:res_scanorigin}

\begin{figure}[h]
    \centering
    \includegraphics[width=0.75\columnwidth]{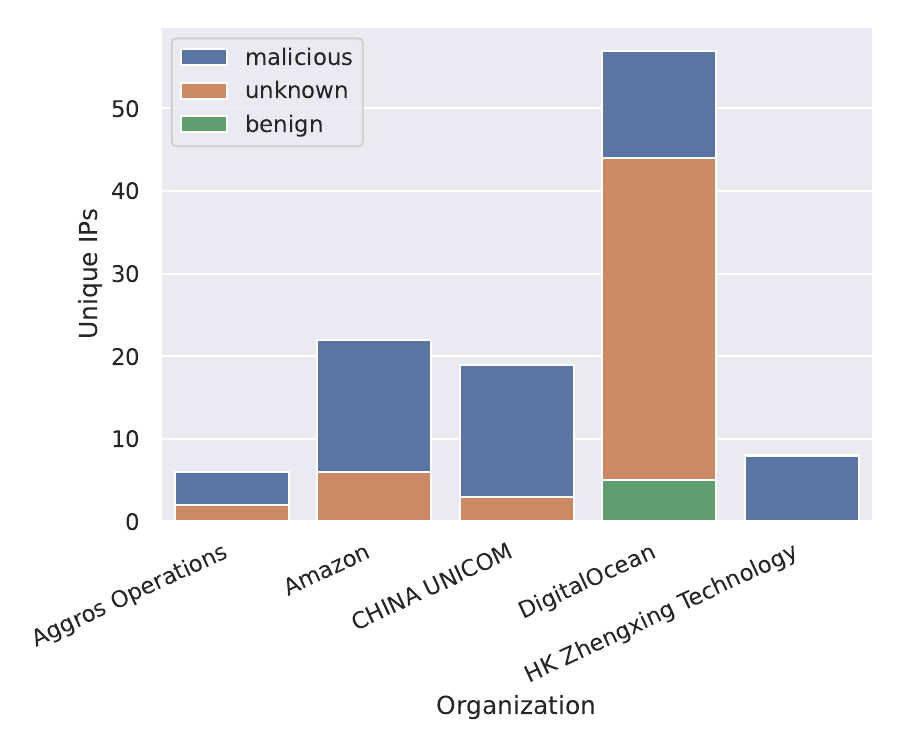}
    \caption{Most frequent organization together with IPs label of Greynoise.}
    \label{fig:org_ips}
\end{figure}

To analyze the presence of malicious actors among the IP sources, we leveraged GreyNoise~\cite{greynoise}. Greynoise is a company that collects, labels, and diagnoses data and provides access to such information to users via API.
The analysis shows that about $51\%$ of unique IP addresses analyzed are labeled in the GreyNoise database as ``malicious''. These actors are generally IP sources flagged as attackers in previous campaigns. The remaining IPs were $23.5\%$ labeled as ``benign'' sources and the remaining part as ``unknown''. ``benign'' refers to research centers or scanning services (e.g., Shodan or Zoomeye) that scan the IP addresses without malicious purposes.
In Figure~\ref{fig:org_ips}, we report the top-5 organizations with the most malicious IP addresses. These organizations are \textit{DigitalOcean, LLC}, \textit{Amazon.com, Inc.}, 
\textit{CHINA UNICOM China169 Backbone}, \textit{Hong Kong Zhengxing Technology Co., Ltd.}, and \textit{
Aggros Operations Ltd.}. Other actors include numerous ISPs from all over the world. 

GreyNoise can also identify the actor related to an IP address, i.e., the entity using the IP address. The actor can differ from the organization to which the address belongs since it commonly happens that ISPs and companies offering cloud computing services or IP addresses block which are used for malicious activities.
All the actors analyzed belong to hosts categorized as ``benign'' and represent legitimate organizations scanning the network to identify exposed and vulnerable services, sometimes also notifying the owners about the dangers they incur. The most frequent benign scanners in both the honeypot include \textit{Stretchoid}, \textit{Censys}, \textit{Bitsight}, \textit{ShadowServer}, \textit{Cortex Xpanse}, and \textit{Shodan}. All the malicious scanner actors are instead unknown by Greynoise except for some IPs belonging to \textit{Stretchoid}, \textit{XMCO.fr}, and \textit{LeakIX}.

\begin{figure}[h]
    \centering
    \includegraphics[width=0.9\columnwidth]{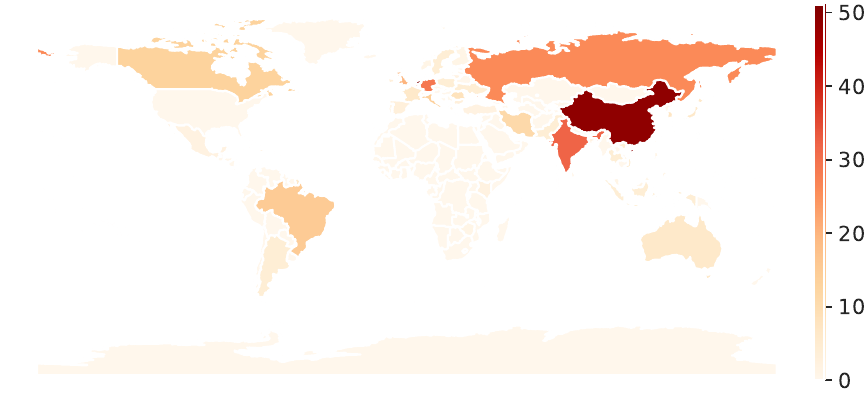}
    \caption{Country of origin of malicious IPs.}
    \label{fig:map_ip}
\end{figure}

In Figure~\ref{fig:map_ip}, we report the country of origin of the unique IPs labeled as malicious. We can note that most of the malicious IPs belong to China, Russia, and India. However, we must note that this may not be the original source of the scan but may be the last hop of a VPN.

\section{Conclusion}\label{sec:conclusion}

This paper presents HoneyEVSE, the first honeypot conceived to emulate a charging column. HoneyEVSE leverages tools like ACN-Sim to simulate the \ac{ev} charging process and Honeyd as a network personality engine. We developed HoneyEVSE to mimic existing \ac{evse} that related work found exposed on the internet, and we performed a data collection 30 days after exposing it. 
Results show that HoneyEVSE obtained a satisfying level of engagement during the data collection, confirming its capability in effectively reproducing a \ac{evse} device. In this direction, future works include the development of more sophisticated \ac{evse} functionalities and services to increase emulation fidelity.

We believe that HoneyEVSE can support the security research on the \ac{v2g} field by enabling novel applications and analysis on \ac{evse} functioning.

\section*{Acknowledgment}
We thank \emph{VSIX}\cite{vsix} for enabling us to install the honeypot and collect data at their \ac{ixp}.

\bibliographystyle{splncs04}
\bibliography{biblio}

\begin{thebibliography}{10}
\providecommand{\url}[1]{\texttt{#1}}
\providecommand{\urlprefix}{URL }
\providecommand{\doi}[1]{https://doi.org/#1}

\bibitem{greynoise}
{GreyNoise Intelligence}, \url{https://greynoise.io/}

\bibitem{openenergymonitor}
Openenergymonitor, \url{https://openenergymonitor.org/}, [Accessed: 05-08-2023]

\bibitem{vsix}
Vsix internet exchange point, \url{https://www.vsix.it/}, [Accessed:
  15-05-2023]

\bibitem{gridpot_proj}
{GridPot github project} (2015), \url{https://github.com/sk4ld/gridpot},
  accessed: 02-05-2023

\bibitem{evse-anatomy}
What is evse? (4 2023), \url{https://ev-lectron.com/blogs/blog/what-is-evse},
  [Accessed: 03-08-2023]

\bibitem{baker2019losing}
Baker, R., Martinovic, I.: Losing the car keys: Wireless phy-layer insecurity
  in ev charging. USENIX (2019)

\bibitem{9522219}
Barbieri, G., Conti, M., Tippenhauer, N.O., Turrin, F.: Assessing the use of
  insecure ics protocols via ixp network traffic analysis. In: 2021
  International Conference on Computer Communications and Networks (ICCCN).
  pp.~1--9 (2021). \doi{10.1109/ICCCN52240.2021.9522219}

\bibitem{brighente2021evscout2}
Brighente, A., Conti, M., Donadel, D., Turrin, F.: Evscout2. 0: electric
  vehicle profiling through charging profile. ACM Transactions on
  Cyber-Physical Systems  (2021)

\bibitem{Buschlinger2019}
Buschlinger, L., Springer, M., Zhdanova, M.: {Plug-and-patch: Secure value
  added services for electric vehicle charging}. ACM International Conference
  Proceeding Series  (2019)

\bibitem{cenys2004development}
Cenys, A., Rainys, D., Radvilavicius, L., Bielko, A.: Development of honeypot
  system emulating functions of database server. Tech. rep., SEMICONDUCTOR
  PHYSICS INST VILNIUS (LITHUANIA) (2004)

\bibitem{EVExchange}
Conti, M., Donadel, D., Poovendran, R., Turrin, F.: Evexchange: A relay attack
  on electric vehicle charging system. In: Atluri, V., Di~Pietro, R., Jensen,
  C.D., Meng, W. (eds.) Computer Security -- ESORICS 2022. pp. 488--508.
  Springer International Publishing, Cham (2022)

\bibitem{conti2022icspot}
Conti, M., Trolese, F., Turrin, F.: Icspot: A high-interaction honeypot for
  industrial control systems. In: 2022 International Symposium on Networks,
  Computers and Communications (ISNCC). pp.~1--4. IEEE (2022)

\bibitem{franco2021survey}
Franco, J., Aris, A., Canberk, B., Uluagac, A.S.: A survey of honeypots and
  honeynets for internet of things, industrial internet of things, and
  cyber-physical systems. IEEE Communications Surveys \& Tutorials
  \textbf{23}(4),  2351--2383 (2021)

\bibitem{garofalaki2022electric}
Garofalaki, Z., Kosmanos, D., Moschoyiannis, S., Kallergis, D., Douligeris, C.:
  Electric vehicle charging: A survey on the security issues and challenges of
  the open charge point protocol (ocpp). IEEE Communications Surveys \&
  Tutorials  (2022)

\bibitem{ISO15118-1}
{Road vehicles — Vehicle-to-Grid Communication Interface — Part 1: General
  information and use-case definition}. Standard, International Organization
  for Standardization, Geneva, CH (Mar 2019)

\bibitem{ISO15118-2}
{Road vehicles — Vehicle-to-Grid Communication Interface — Part 2: Network
  and application protocol requirements}. Standard, International Organization
  for Standardization, Geneva, CH (Mar 2014)

\bibitem{ACNPortal}
Johansson, D., Lee, Z.J., Sharma, S.: Acn portal (06 2021),
  \url{https://github.com/zach401/acnportal}

\bibitem{kohler2022brokenwire}
K{\"o}hler, S., Baker, R., Strohmeier, M., Martinovic, I.: Brokenwire: Wireless
  disruption of ccs electric vehicle charging. arXiv preprint arXiv:2202.02104
  (2022)

\bibitem{ACNSim}
Lee, Z., Sharma, S., Johansson, D., Low, S.: Acn-sim: An open-source simulator
  for data-driven electric vehicle charging research. IEEE Transactions on
  Smart Grid  \textbf{PP} (12 2020). \doi{10.1109/TSG.2021.3103156}

\bibitem{lee2019acn}
Lee, Z.J., Li, T., Low, S.H.: Acn-data: Analysis and applications of an open ev
  charging dataset. In: Proceedings of the Tenth ACM International Conference
  on Future Energy Systems. pp. 139--149 (2019)

\bibitem{luo2017iotcandyjar}
Luo, T., Xu, Z., Jin, X., Jia, Y., Ouyang, X.: Iotcandyjar: Towards an
  intelligent-interaction honeypot for iot devices. Black Hat  \textbf{2017},
  1--11 (2017)

\bibitem{HoneyPLC}
López-Morales, E., Rubio, C., Doupé, A., Shoshitaishvili, Y., Bao, T., Ahn,
  G.J.: Honeyplc: A next-generation honeypot for industrial control systems.
  pp. 279--291 (10 2020). \doi{10.1145/3372297.3423356}

\bibitem{mashima2019s}
Mashima, D., Li, Y., Chen, B.: Who's scanning our smart grid? empirical study
  on honeypot data. In: 2019 IEEE Global Communications Conference (GLOBECOM).
  pp.~1--6. IEEE (2019)

\bibitem{shodan_guide}
Matherly, J.: {Complete Guide to Shodan},
  \url{https://ia800705.us.archive.org/17/items/shodan-book-extras/shodan/shodan.pdf}

\bibitem{Shodan}
Matherly, J.: Complete guide to shodan (10 2016),
  \url{https://ia800705.us.archive.org/17/items/shodan-book-extras/shodan/shodan.pdf}

\bibitem{murillo2018virtual}
Murillo, A.F., C{\'o}mbita, L.F., Gonzalez, A.C., Rueda, S., Cardenas, A.A.,
  Quijano, N.: A virtual environment for industrial control systems: A
  nonlinear use-case in attack detection, identification, and response. In:
  Proceedings of the 4th Annual Industrial Control System Security Workshop.
  pp. 25--32 (2018)

\bibitem{nasrchargeprint}
Nasr, T., Torabi, S., Bou-Harb, E., Fachkha, C., Assi, C.: Chargeprint: A
  framework for internet-scale discovery and security analysis of ev charging
  management systems. In: NDSS (2023)

\bibitem{9903648}
Panda, S., Rass, S., Moschoyiannis, S., Liang, K., Loukas, G., Panaousis, E.:
  Honeycar: A framework to configure honeypot vulnerabilities on the internet
  of vehicles. IEEE Access  \textbf{10},  104671--104685 (2022).
  \doi{10.1109/ACCESS.2022.3210117}

\bibitem{petre2019honeypot}
Petre, C.A., Korodi, A.: Honeypot inside an opc ua wrapper for water pumping
  stations. In: 2019 22nd International Conference on Control Systems and
  Computer Science (CSCS). pp. 72--77. IEEE (2019)

\bibitem{Honeyd}
Provos, N.: Honeyd: A virtual honeypot daemon (extended abstract)  (01 2003)

\bibitem{rahmatullah2016implementation}
Rahmatullah, D.K., Nasution, S.M., Azmi, F.: Implementation of low interaction
  web server honeypot using cubieboard. In: 2016 International Conference on
  Control, Electronics, Renewable Energy and Communications (ICCEREC). pp.
  127--131. IEEE (2016)

\bibitem{sharma2018survey}
Sharma, S., Kaul, A.: A survey on intrusion detection systems and honeypot
  based proactive security mechanisms in vanets and vanet cloud. Vehicular
  communications  \textbf{12},  138--164 (2018)

\bibitem{statista_ev}
Statista: Electric vehicles - worldwide (2023),
  \url{https://www.statista.com/outlook/mmo/electric-vehicles/worldwide},
  [Online: accessed Apr-2023]

\end{thebibliography}

\end{document}